\documentclass[12pt,letterpaper]{article}
\usepackage{amssymb}

\usepackage{graphicx}
\usepackage{amsmath}

\input{tcilatex}

\begin{document}

\author{V. V. Prosentsov\thanks{%
e-mail: prosentsov@yahoo.com} \\
Stationsstraat 86, 5751 HH, Deurne, The Netherlands}
\title{Resonance scattering of light by spherical photonic clusters: analytical
approach\thanks{%
The paper was submitted to JOSA\ B.}}
\maketitle

\begin{abstract}
Scattering of light by the photonic clusters made of small particles is
studied with the help of the quasiperiodicity condition and the local
perturbation method. The analytical expression for the field scattered by
the cluster is presented and the conditions of resonance scattering are
found. The conditions of the zero scattering by the cluster are also found.
\end{abstract}

\section{Introduction}

Light scattering is a classical problem of electrodynamics which with
emergency of photonic crystals became actual topic with many exiting
applications \cite{Lavers}, \cite{Heath}. The scattering by photonic
crystals is difficult to treat analytically because of they complex
structure and large number of particles or cells involved in calculations.
In practice, real photonic crystals are clusters with finite dimensions and
finite number of particles (or cells). In many theoretical works the
clusters were approximated by infinite crystals and electromagnetic fields
inside the clusters assumed to be quasiperiodic (see for example works \cite
{Pavarini}, \cite{Wang} and references therein). Even in experimental works
the same technic was used successfully to explain observed results \cite
{Ivans}. Alternatively, the light scattering by photonic clusters was
studied rigorously by using the finite difference time domain method \cite
{Kole} and the local perturbation method \cite{VpAd1}, \cite{VpAd2} (often
called also discrete dipole approximation or coupled dipole approximation
method \cite{Chaumet}). These numerical methods require number crunching
machines with substantial computational power and memory. The principal
limitation of the numerical methods is that final answer is a ''naked''
number without a functional dependence. Some kind of trend can be obtained
after several computations, functional dependencies however, can not be
predicted. In the local perturbation method, for example, the solution of
the large system of linear equation is required and final result includes
sums over all particles \cite{VpAd1}. Analytical solution can be obtained in
principle, however it will be not transparent for understanding. Under
certain conditions (see for example work \cite{VpAd2}), the cluster can be
treated as infinite crystal where the quasiperiodicity exists. Due to the
quasiperiodicity, the field should be calculated in one point only. This
drastically simplifies calculations in the local perturbation method and the
final results become more transparent. Firstly, it was done for
one-dimensional infinite crystals \cite{Markel} and afterwards it was
generalized to three-dimensional structures \cite{Draine}.

In our work we studied the light scattering from the spherical cluster by
using both the local perturbation method and the quasiperiodicity. In
contrast to previous works we obtained much simpler analytical expressions
for the fields and analyzed resonance scattering. The resonance frequencies
of the scattered field were calculated and resonance width was discussed. It
was found that the field scattered by the spherical cluster has ''cloaking''
frequencies at which the field is absent.

\section{The formalism}

The formalism we use is presented in many works (see for example \cite{VpAd1}%
-\cite{Chaumet1}) and we only briefly present it here for convenience and
consistency. Consider the photonic cluster made of identical particles which
characteristic size $L$ is small compared to incident wavelength $\lambda
_{0}$. The electric field $\mathbf{E}$\ propagating in the host medium
filled with the $N$ particles is described by the following equation \cite
{BassVpFr}

\begin{equation}
\left( \bigtriangleup -\mathbf{\nabla }\otimes \mathbf{\nabla }%
+k_{0}^{2}\right) \mathbf{E}(\mathbf{r})+k_{0}^{2}\frac{(\varepsilon
_{sc}-\varepsilon _{0})}{\varepsilon _{0}}\sum_{n=0}^{N-1}f(\mathbf{r}-%
\mathbf{r}_{n})\mathbf{E}(\mathbf{r}_{n})=\mathbf{S}(\mathbf{r}),
\label{ps11}
\end{equation}
where

\begin{equation*}
\mathbf{\;}k_{0}=\left| \mathbf{k}_{0}\right| =\frac{2\pi }{\lambda _{0}}=%
\frac{\omega }{c}\sqrt{\varepsilon _{0}},\;f(\mathbf{r}-\mathbf{r}%
_{n})=\left\{ 
\begin{array}{cc}
1\text{,} & \text{inside particle } \\ 
0\text{,} & \text{outside particle}
\end{array}
\right. .
\end{equation*}
Here $\bigtriangleup $ is Laplacian and $\mathbf{\nabla }$\ nabla operators, 
$\otimes $ defines tensor product, $k_{0}$ is a wave number in the host
medium ($\omega $ is the angular frequency and $c$ is the light velocity in
vacuum), $\varepsilon _{sc}$ and $\varepsilon _{0}$ are the permittivity of
the particles and the medium respectively, $f$ is the function describing
the shape of the scatterers, and $\mathbf{S}$ is the field source. Note,
that the equation (\ref{ps11}) is an approximate one and this is true only
when the small scatterers ($k_{0}L\ll 1$) are considered. This equation is
already solvable, however it will invoke solution of $N$ by $N$ system of
linear equations for each frequency $\omega $. In addition, the final result
will include summation over all $N$ particles that reduces solution
transparency even further. The solution of the Eq. (\ref{ps11}) can be
significantly simplified when the field $\mathbf{E}$ is quasiperiodic, i.e.
when 
\begin{equation}
\mathbf{E}(\mathbf{r+r}_{n})=\mathbf{E}(\mathbf{r})e^{i\mathbf{k}_{0}\mathbf{%
\cdot r}_{n}}.  \label{ps11a}
\end{equation}
We note that all particles are supposed to be identical for the condition (%
\ref{ps11a}) to be valid. By using relation (\ref{ps11a}) the equation (\ref
{ps11}) can be written in the following form

\begin{equation}
\mathbf{E}(\mathbf{r})=\mathbf{E}_{0}(\mathbf{r})+k_{0}^{2}\frac{%
(\varepsilon _{sc}-\varepsilon _{0})}{\varepsilon _{0}}\left( \widehat{I}+%
\frac{\mathbf{\nabla }\otimes \mathbf{\nabla }}{k_{0}^{2}}\right) \mathbf{E}%
(0)\sum_{n=0}^{N-1}e^{i\mathbf{k}_{0}\mathbf{\cdot r}_{n}}\Phi _{n}(\mathbf{r%
}),  \label{ps12}
\end{equation}
where

\begin{eqnarray}
\mathbf{E}(0) &=&\left. \mathbf{E}(\mathbf{r}_{0})\right| _{\mathbf{r}%
_{0}=0},\;\mathbf{r}_{n}\equiv \sum_{i=1}^{3}m_{i}^{n}\mathbf{d}_{i},
\label{ps12a} \\
\Phi _{n}(\mathbf{r}) &=&\int_{-\infty }^{\infty }\frac{\widetilde{f}(%
\mathbf{q})e^{i\mathbf{q\cdot (r-r}_{n})}}{(q^{2}-k_{0}^{2})}d\mathbf{q,\;}\;%
\widetilde{f}(\mathbf{q})=\frac{1}{8\pi ^{3}}\int_{-\infty }^{\infty }f(%
\mathbf{r})e^{-i\mathbf{q\cdot r}}d\mathbf{r.}  \label{ps12b}
\end{eqnarray}
Here $\widehat{I}$ is the $3\times 3$ unitary tensor in polarization space
and $\mathbf{r}_{n}$ $\equiv \sum_{i=1}^{3}m_{i}^{n}\mathbf{d}_{i}$ is the
radius vector of the n-th particle, where $m_{i}^{n}$ are integers ($%
m_{i}^{0}=0$) and $\mathbf{d}_{i}$ are the basis vectors of the crystal. The
field $\mathbf{E}(0)$ is the field at the $0$-th particle and $\widetilde{f}$
is the Fourier transform of the function $f$. The incident field $\mathbf{E}%
_{0}$ is created by the source $\mathbf{S}$ in the host medium and it is not
important for our consideration (the expression for $\mathbf{E}_{0}$ is
presented by formula (\ref{ap1}) in Appendix). The formula (\ref{ps12}) is
rather general one and it describes the field in the medium with photonic
cluster of arbitrary form made of small particles of arbitrary form. The
formula shows that the total field $\mathbf{E}$ is a sum of the incident
field $\mathbf{E}_{0}$ propagating in the medium and the field due to the
scattering by the cluster formed by $N$ particles. The interference between
the scatterers as well as their resonance properties are taken into account
by only one term $\mathbf{E}(0)$ that should be found by solving the system
of $9$ linear equations obtained by substituting $\mathbf{r}=\mathbf{r}%
_{0}=0 $ into (\ref{ps12}). The formula (\ref{ps12}) can be simplified even
further when the distance between the observer and an $n$-th scatterer ($%
R_{n}$) is large, i.e. when $R_{n}\gg L$. In this case the integral $\Phi
_{n}$ can be calculated approximately. We note also that the integral $\Phi
_{n}$ can be calculated exactly at least for the spherical particles. When $%
R_{n}\gg L$ integration in Eq. (\ref{ps12}) gives 
\begin{equation}
\mathbf{E}(\mathbf{r})=\mathbf{E}_{0}(\mathbf{r})+k_{0}^{2}V\frac{%
(\varepsilon _{sc}-\varepsilon _{0})}{4\pi \varepsilon _{0}}\left( \widehat{I%
}+\frac{\mathbf{\nabla }\otimes \mathbf{\nabla }}{k_{0}^{2}}\right) \mathbf{E%
}(0)\sum_{n=0}^{N-1}e^{i\mathbf{k}_{0}\mathbf{\cdot r}_{n}}\frac{%
e^{ik_{0}R_{n}}}{R_{n}},  \label{ps14}
\end{equation}
where

\begin{equation}
R_{n}=\left| \mathbf{r}-\mathbf{r}_{n}\right| \gg L.  \label{ps15}
\end{equation}
Here $R_{n}$ is the distance between the observation point and $n$-th
scatterer and $V$ is the scatterer's volume.

It is interesting to note that the expression (\ref{ps12}) describes the
field inside infinite photonic crystal when $N\rightarrow \infty $. In
addition, the formula explicitly shows that even small losses in the host
medium (small positive imaginary part of $k_{0}$) limit the number of
interacting particles in the infinite photonic crystal.

The formulae (\ref{ps12}) and (\ref{ps14}) can be used with two
requirements: the quasiperiodicity and small size of the particles. The
second requirement can be overcome by further subdivision of the particles.
When the scatterers are not sufficiently small ($k_{0}L\geq 0.2$, for
example), they can be subdivided into smaller particles with size $L_{sub}$
such that $k_{0}L_{sub}\leq 0.1$.

It should be emphasized here that the main assumption of the work is the
existence of the quasiperiodicity (expression (\ref{ps11a})) for a finite
size photonic system. Obviously, the assumption is not correct for the
lossless systems where light freely propagates from one boundary to another.
Weak interference between the scatterers (due to small optical contrast or
large distance, for example) and small losses allow to treat finite system
as infinite one because of the effective absence of the boundaries. In this
case the expression (\ref{ps11a}) should be correct.

\subsection{Fields in scalar approximation}

In this subsection the results obtained in scalar approximation are
presented. These results are valid when light depolarization is negligible.
We note that the results are important for general understanding and
comparison with vector case. For definiteness we assume that the cluster is
made of spherical particles with radius $L$ (for particles with other shapes
the similar results can be obtained). In accordance with the formula (\ref
{ps14}), the field in scalar approximation is

\begin{equation}
E(\mathbf{r})=E_{0}(\mathbf{r})+k_{0}^{2}L^{3}\frac{(\varepsilon
_{sc}-\varepsilon _{0})}{3\varepsilon _{0}}E(0)\sum_{n=0}^{N-1}e^{i\mathbf{k}%
_{0}\mathbf{\cdot r}_{n}}\frac{e^{ik_{0}R_{n}}}{R_{n}},  \label{ps21}
\end{equation}
where

\begin{equation}
E(0)=\frac{E_{0}(0)}{D}  \label{ps22}
\end{equation}
and

\begin{eqnarray}
D &=&1-\alpha -k_{0}^{2}L^{3}\frac{(\varepsilon _{sc}-\varepsilon _{0})}{%
3\varepsilon _{0}}\sum_{n=1}^{N-1}e^{i\mathbf{k}_{0}\mathbf{\cdot r}_{n}}%
\frac{e^{ik_{0}r_{n}}}{r_{n}}  \label{ps23} \\
\alpha &=&k_{0}^{2}L^{2}\frac{(\varepsilon _{sc}-\varepsilon _{0})}{%
2\varepsilon _{0}}\left( 1+i2k_{0}L/3\right) ,\;r_{n}=\left| \mathbf{r}%
_{n}\right| .
\end{eqnarray}
The expressions (\ref{ps22}) and (\ref{ps23})\ already show that resonance
properties of the field $E(0)$ are specified by the parameters of the $0$-th
particle via $\alpha $\ and by the rest of the cluster via the sum in Eq. (%
\ref{ps23}). The contribution of the rest of the cluster is controlled by
its structure and varies from significant to negligible (as shown in Eqs. (%
\ref{ps25}) and (\ref{ps29}) below). The sums in the expressions (\ref{ps21}%
) and (\ref{ps23}) can be calculated numerically, however the final result
will be not transparent for analysis. Alternatively, these sums can be
evaluated analytically by using the integration (see formulae (\ref{ap2})
and (\ref{ap3}) in Appendix) when $k_{0}d_{i}\ll 1$. This condition is
extremely convenient for our consideration, because for close packed
clusters (when $d_{i}=2L$) it coincides with the assumption required for the
method we use, i.e. $k_{0}L\ll 1$. For the cluster it is naturally to use
the condition $r\gg r_{n}$ (observer is far from the cluster). The final
expression for the field in the medium with the spherical cluster is

\begin{equation}
E(\mathbf{r})=E_{0}(\mathbf{r})+k_{0}^{2}L^{3}\beta \frac{(\varepsilon
_{sc}-\varepsilon _{0})}{3\varepsilon _{0}}E(0)\frac{e^{ik_{0}r}}{r},
\label{ps25}
\end{equation}
where

\begin{eqnarray}
\beta &=&1-\frac{\pi }{6}+\frac{4\pi }{k_{sc}^{3}d^{3}}\left[ \sin (\zeta
)-\zeta \cos (\zeta )\right] ,  \label{ps25a} \\
k_{sc} &=&\left| \mathbf{k}_{sc}\right| =k_{0}\mu ,\;\mu =\left| \frac{%
\mathbf{k}_{0}}{k_{0}}-\frac{\mathbf{r}}{r}\right| ,\;\zeta =k_{sc}(\Lambda
+d/2).  \notag
\end{eqnarray}
The field $E(0)$ is described by the expression (\ref{ps22}) where

\begin{eqnarray}
D &=&1-\alpha -2\pi \frac{(\varepsilon _{sc}-\varepsilon _{0})}{3\varepsilon
_{0}}\frac{L^{3}}{d^{3}}\left( 
\begin{array}{c}
k_{0}^{2}d^{2}/4-\sin ^{2}(\xi )+ik_{0}d \\ 
+i\sin (2\xi )/2-i\xi
\end{array}
\right) ,  \label{ps26} \\
\xi &=&k_{0}(\Lambda +d/2).
\end{eqnarray}
The expression (\ref{ps25}) is the field in the medium filled with small
spherical particles forming spherical cluster of radius $\Lambda $. The
particles are positioned in cubic lattice with period $d$. The formulae (\ref
{ps25})-(\ref{ps26}) is the main result of this subsection and we will
discuss them thoroughly.

Consider first the scattered field. The formulae (\ref{ps25}) and (\ref
{ps25a}) show that the scattered field oscillates as function of the
scattering wave number $k_{sc}$\ and the radius of the cluster $\Lambda $.
The expressions (\ref{ps25}) and (\ref{ps25a}) show also influence of the
scattering vector $\mathbf{k}_{sc}$. When $\mathbf{k}_{sc}=0$ (backward
scattering) $\beta \sim N$ and the field is proportional to the cluster
volume $NL^{3}$. The formula (\ref{ps25}) also shows that the scattering
field is zero when $\beta =0$, i.e. when

\begin{equation}
\zeta \cos (\zeta )-\sin (\zeta )=\frac{k_{sc}^{3}d^{3}}{4\pi }\left( 1-%
\frac{\pi }{6}\right) .  \label{ps29}
\end{equation}
Taking into account that $k_{sc}^{3}d^{3}\ll 1$, the solution $\zeta _{0}$
of the equation (\ref{ps29}) can be found in the following form

\begin{equation}
\zeta _{0}=\pi \left( M+\frac{1}{2}\right) +\frac{1}{\pi \left( M+1/2\right) 
}=k_{sc}(\Lambda +d/2),  \label{ps29a}
\end{equation}
where $M$ is a positive integer. The existence of the solution means that
the cluster is actually invisible (cloaking effect) at the ''cloaking''
frequency $\omega _{cl}$ which is

\begin{equation}
\omega _{cl}=\frac{c}{\sqrt{\varepsilon _{0}}}\frac{\zeta _{0}}{\mu (\Lambda
+d/2)}.  \label{ps29b}
\end{equation}
We note here that the expression (\ref{ps29b})\ explicitly shows that the
cloaking frequencies of the photonic cluster depend on the cluster radius $%
\Lambda $ and the direction of the scattering wave vector (defined by $\mu $%
).

Consider now the resonance properties of the scattered field. The importance
of the Eq. (\ref{ps26}) is that Re$(D)=0$ defines the resonance frequencies
of the field scattered by the cluster. For the dielectric scatterers the
equation for the resonance frequencies can be expressed in the form

\begin{equation}
\frac{2\varepsilon _{0}}{(\varepsilon _{sc}-\varepsilon _{0})}%
=k_{0}^{2}L^{2}\left( 1-\frac{\pi L}{6d}\right) +\frac{4\pi }{3}\frac{L^{3}}{%
d^{3}}\sin ^{2}(\xi ).  \label{ps26c}
\end{equation}
Note that when the period $d$\ of the cluster increases, the expression (\ref
{ps26c}) reproduces equation for the resonance frequencies of the single
particle. Consider two most interesting cases.

I. When $\sin (\xi )=0$ the solution of the Eq. (\ref{ps26c}) is

\begin{equation}
\omega _{r1}=\frac{\sqrt{2}c}{L\sqrt{\varepsilon _{sc}-\varepsilon _{0}}}%
\left( 1-\frac{\pi }{6}\frac{L}{d}\right) ^{-1/2}.  \label{ps26d}
\end{equation}
The formula (\ref{ps26d}) clearly shows that the resonance frequency of the
single particle is the principal contribution and the period of the cluster
only tunes the resonance frequency of the scattered field (keep in mind that 
$d\geq 2L$). In addition, the resonance frequency is defined from the
condition $\sin (\xi )=0$ and it is

\begin{equation}
\omega _{r2}=\frac{\pi M}{\sqrt{\varepsilon _{0}}}\frac{c}{(\Lambda +d/2)}
\label{ps 27}
\end{equation}
where $M$ is a positive integer.

II. When $\frac{L}{d}\sin (\xi )>k_{0}^{2}d^{2}$ the solution of the Eq. (%
\ref{ps26c}) is

\begin{equation}
\omega _{r}=\frac{c}{\sqrt{\varepsilon _{0}}}\frac{\left| \arcsin \left( \pm 
\sqrt{\frac{3d^{3}}{2\pi L^{3}}\frac{\varepsilon _{0}}{(\varepsilon
_{sc}-\varepsilon _{0})}}\right) \right| }{(\Lambda +d/2)}.  \label{ps27a}
\end{equation}
The formula (\ref{ps27a}) explicitly shows that the resonance frequency of
the field scattered by the cluster is periodic and rather complex function
of the cluster parameters (period $d$, scatterer size $L$, optical contrast $%
\varepsilon _{sc}-\varepsilon _{0}$) and the frequency is inversely
proportional to the cluster size $\Lambda $. We note here that the resonance
will occur when the following condition is satisfied

\begin{equation}
0<\frac{3d^{3}}{2\pi L^{3}}\frac{\varepsilon _{0}}{(\varepsilon
_{sc}-\varepsilon _{0})}\leq 1.  \label{ps27bc}
\end{equation}
The condition (\ref{ps27bc}) shows that for close packed cluster (when $d=2L$%
), for example, the condition $\varepsilon _{sc}\geq 5\varepsilon _{0}$
should be satisfied.

Finally, we consider the resonance width $\gamma $ which near the resonance
(when $\omega =\omega _{r}$) is defined in the following way

\begin{equation}
\gamma (\omega )=\frac{\left| \text{Im}(D(\omega ))\right| }{\left| \frac{%
\partial \text{Re}(D)}{\partial \omega }\right| _{\omega =\omega _{r}}}.
\label{ps28}
\end{equation}
The resonance is sharp when $\gamma (\omega _{r})\ll \omega _{r}$. In
accordance with the formula (\ref{ps28}) and (\ref{ps26}) the resonance
width can be presented in the form

\begin{equation}
\gamma (\omega )\simeq \frac{\omega \left| \frac{k_{0}L}{3}+\frac{2L}{d}%
\frac{1}{k_{0}^{2}d^{2}}\left( \xi -\frac{\sin (2\xi )}{2}\right) \right| }{%
\left| 1-\frac{L}{d}+\frac{2L}{d}\frac{\xi }{k_{0}^{2}d^{2}}\sin (2\xi
)\right| }.  \label{ps28a}
\end{equation}
The expression (\ref{ps28a}) explicitly shows that the resonance width is
sensitive to the cluster size $\Lambda \sim dN^{1/3}$, period $d$, and
characteristic size of the particles $L$. The formula (\ref{ps28a}) also
confirms that the resonance width of the field scattered by the cluster with
extremely large period ($k_{0}^{2}Ld\gg \frac{L}{d}N^{1/3}$) coincides with
the resonance width of the field scattered by the single particle. When the
cluster is relatively dense ($d\sim 2L$) and $\xi \sin (2\xi )\gtrsim 1$,
the resonance width takes the form

\begin{equation}
\gamma (\omega _{r})\simeq \frac{\omega _{r}}{\left| \sin (2\xi )\right| }%
\gg \omega _{r},  \label{ps28b}
\end{equation}
i.e. the resonance of such cluster is not sharp.

\subsection{Fields in vector case}

The vector case will be considered in a similar manner as the scalar one.
Without the loss of generality we can assume that the wave vector of the
incident wave $\mathbf{k}_{0}$ is directed along $z$ direction. This
assumption simplifies the calculations of the sums in Eq. (\ref{ps14}).
After summation in formula (\ref{ps14}) the field in the far zone ($%
k_{0}r\gg 1$) can be presented in the following form

\begin{eqnarray}
\mathbf{E}(\mathbf{r}) &=&\mathbf{E}_{0}(\mathbf{r})+k_{0}^{2}L^{3}\beta 
\frac{(\varepsilon _{sc}-\varepsilon _{0})}{3\varepsilon _{0}}\frac{%
e^{ik_{0}r}}{r}\left( \widehat{I}-\mathbf{l}\otimes \mathbf{l}\right) 
\mathbf{E}(0),  \label{ps31} \\
\mathbf{l} &=&\mathbf{r}/r  \notag
\end{eqnarray}
where the parameter $\beta $ is given by the formula (\ref{ps25a}) and the
components of the field $E_{j}(0)$ are

\begin{equation}
E_{j}(0)=\frac{E_{0,j}(0)}{D_{j}},\;(j=x,y,z).  \label{ps31a}
\end{equation}
Here the denominator $D_{j}$ is described by the following formulae

\begin{eqnarray}
D_{x,y} &=&1-\chi -\frac{(\varepsilon _{sc}-\varepsilon _{0})}{3\varepsilon
_{0}}\frac{4\pi L^{3}}{d^{3}}\left. \left( 
\begin{array}{c}
\frac{\allowbreak e^{2ix}}{2}\left( -\frac{1}{2}-\frac{i}{x}+\frac{2}{x^{2}}+%
\frac{i}{x^{3}}\right) \\ 
+\frac{i}{2}\left( x-\frac{1}{x}-\allowbreak \frac{1}{x^{3}}\right)
\end{array}
\right) \right| _{x=k_{0}d/2}^{x=\xi }  \label{ps32} \\
D_{z} &=&1-\chi -\frac{(\varepsilon _{sc}-\varepsilon _{0})}{3\varepsilon
_{0}}\frac{4\pi L^{3}}{d^{3}}\left. \left( 
\begin{array}{c}
\allowbreak e^{2ix}\left( \frac{i}{x}-\frac{2}{x^{2}}-\frac{i}{x^{3}}\right)
\\ 
+\frac{i}{x}+\allowbreak \frac{i}{x^{3}}
\end{array}
\right) \right| _{x=k_{0}d/2}^{x=\xi }.  \notag
\end{eqnarray}
When the dimension of the cluster is relatively large ($\xi \gg 1$), the
formulae (\ref{ps32}) can be simplified and written in the following form

\begin{eqnarray}
D_{j} &=&1-\chi -\frac{(\varepsilon _{sc}-\varepsilon _{0})}{3\varepsilon
_{0}}\frac{4\pi L^{3}}{d^{3}}\vartheta _{j},  \label{ps32a} \\
\chi &=&\frac{(\varepsilon _{sc}-\varepsilon _{0})}{(-3)\varepsilon _{0}}%
\left( 1-k_{0}^{2}L^{2}-i\frac{2}{3}k_{0}^{3}L^{3}\right) ,\;\xi
=k_{0}(\Lambda +d/2),  \notag
\end{eqnarray}
where

\begin{equation}
\vartheta _{j}=\left\{ 
\begin{array}{cc}
i\xi /2-\allowbreak e^{2i\xi }/4+7/12, & j=x,y \\ 
i(1+\allowbreak e^{2i\xi })/\xi -2/3, & j=z
\end{array}
\right. .  \label{ps32b}
\end{equation}
As well as in scalar case the scattered field oscillates with the cluster's
size and vanishes when $\beta =0$. In addition, the influence of the
observation position is important via the scattering wave number $k_{sc}$
and the depolarization term $\widehat{I}-\mathbf{l}\otimes \mathbf{l}$.

The equations for the resonance frequencies are defined from the condition Re%
$(D_{j})=0$. The expressions (\ref{ps32a}) and (\ref{ps32b}) show that the
imaginary part of $D_{x,y}$ grows with the size of the cluster $\Lambda $
preventing a sharp resonance. At the same time, the imaginary part of $D_{z}$
decreases with the size of the cluster making the sharp resonance possible.
Below we will study only the sharp resonance. The equation for the resonance
frequencies is defined from the condition Re$(D_{z})=0$ and the equation has
the following form

\begin{equation}
\frac{\varepsilon _{sc}+2\varepsilon _{0}}{(\varepsilon _{sc}-\varepsilon
_{0})}=k_{0}^{2}L^{2}-\frac{8\pi L^{3}}{3d^{3}}.  \label{ps33}
\end{equation}
In this case the resonance frequency is

\begin{equation}
\omega _{r}=\frac{c}{\sqrt{\varepsilon _{0}}L}\sqrt{\frac{\varepsilon
_{sc}+2\varepsilon _{0}}{(\varepsilon _{sc}-\varepsilon _{0})}+\frac{8\pi
L^{3}}{3d^{3}}}.  \label{ps35}
\end{equation}
The solution (\ref{ps35}) shows that the resonance frequency of the
relatively large cluster (when the field is quasiperiodic) is actually
modified resonance frequency of the single scatterer. When the period\ $d$
of the cluster grows, the modification tends to zero very fast. Note that
for the close packed clusters ($d\sim 2L$) the modification is essential and
can not be neglected.

It should be emphasized that the expression (\ref{ps35}) is very important
and elegant one: it describes the resonance for the finite structure and
does not contain the size of the structure. This effect can be explained by
the usage of the quasiperiodic condition suitable for infinite structures.

\section{Conclusions}

The analytical expressions for the field scattered by the spherical photonic
cluster made of small spherical particles have been presented. The scattered
field and its resonance properties have been studied. The resonance
frequencies of the cluster were calculated. It has been shown that the
resonance frequency of the scattered field is a function of the cluster's
period and the characteristic size of the scatterers.

The ''cloaking'' frequencies\ at which the cluster does not scatter have
been found and they have been calculated explicitly.

\section{Acknowledgment}

I thank my wife Lucy for moral support and understanding.

\section{Appendix}

\subsection{The electric field created by a source}

The electric field $\mathbf{E}_{0}$ created by the source $\mathbf{S}$ in
vacuum can be found by using the following formula (see for example \cite
{Jackson})

\begin{equation}
\mathbf{E}_{0}(\mathbf{r})=\int_{-\infty }^{\infty }\widehat{G}(\mathbf{r},%
\mathbf{r}^{\prime })\mathbf{S}(\mathbf{r}^{\prime })d\mathbf{r}^{\prime }, 
\tag{A1}  \label{ap1}
\end{equation}
where $\widehat{G}(\mathbf{r},\mathbf{r}^{\prime })$ is the vacuum Green's
tensor which is

\begin{equation}
\widehat{G}(\mathbf{r},\mathbf{r}^{\prime })=-\left( \widehat{I}+\frac{%
\mathbf{\nabla }\otimes \mathbf{\nabla }}{k_{0}^{2}}\right) \frac{e^{i%
\mathbf{k}_{0}\mathbf{\cdot }\left| \mathbf{r}-\mathbf{r}^{\prime }\right| }%
}{4\pi \left| \mathbf{r}-\mathbf{r}^{\prime }\right| }.  \tag{A1.1}
\label{ap11}
\end{equation}

\subsection{The series summation via integration}

Consider the sum 
\begin{equation}
\sigma =\sum_{n=1}^{N-1}e^{ik_{0}x_{n}},\;(x_{n}=nd),  \tag{A2}  \label{ap2}
\end{equation}
where $n$ is an integer and $d$ is the period for particles placed at $x_{n}$%
. When $k_{0}d\ll 1$ (long wavelength approximation), the function $%
e^{ik_{0}x_{n}}$ varies slowly and the sum (\ref{ap2}) can be replaced by
integration in the following way

\begin{equation}
\sigma =\sum_{n=1}^{N-1}e^{ik_{0}x_{n}}\frac{\delta x_{n}}{\delta x_{n}}%
\simeq \frac{1}{d}\int_{d/2}^{N-d/2}e^{ik_{0}x}dx.  \tag{A3}  \label{ap3}
\end{equation}
Here we used the smallest space bin $\delta x_{n}=d$.

\end{document}